\documentclass[submission,copyright,creativecommons]{eptcs}
\providecommand{\event}{ACL2 Workshop 2017} 
\usepackage{amsmath,amssymb,amsfonts,amsthm}
\usepackage{fancyvrb}
\usepackage[shortcuts]{extdash}
\usepackage{color}
\usepackage{framed}
\usepackage{xspace}
\usepackage{listings}
\newcommand{\becomes}{`becomes'\xspace}

\title{A Versatile, Sound Tool for Simplifying Definitions}
\author{Alessandro Coglio
\institute{Kestrel Institute\\
3260 Hillview Avenue, Palo Alto, CA 94304, USA}
\email{coglio@kestrel.edu}
\and
Matt Kaufmann
\institute{Department of Computer Science\\
The University of Texas at Austin, Austin, TX, USA}
\email{kaufmann@cs.utexas.edu}
\and
Eric W. Smith
\institute{Kestrel Institute\\
3260 Hillview Avenue, Palo Alto, CA 94304, USA}
\email{eric.smith@kestrel.edu}
}
\def\titlerunning{Simplify-defun}
\def\authorrunning{A. Coglio, M. Kaufmann \& E. W. Smith}
\begin{document}
\maketitle

\begin{abstract}

  We present a tool, {\tt simplify-defun}, that transforms the
  definition of a given function into a simplified definition of a new
  function, providing a proof checked by ACL2 that the old and new functions are equivalent.
  When appropriate it also generates termination and
  guard proofs for the new function.  We explain how the tool is
  engineered so that these proofs will succeed.  Examples illustrate
  its utility, in particular for program transformation
  in synthesis and verification.

\end{abstract}

\section{Introduction}
\label{sec:intro}
We present a tool, {\tt simplify-defun}, that transforms the
definition of a given function into a simplified definition of a new
function, providing a proof that the old and new functions are equivalent.
When appropriate it also generates termination and
guard proofs for the new function.  Since the generated proofs are submitted to
ACL2, {\tt simplify-defun} need not be trusted: its soundness only depends on
the soundness of ACL2. The new function is a `simplified' version of the
original function in much the same sense that ACL2 `simplifies' terms during
proofs --- via rewrite, type-set, forward chaining, and linear arithmetic rules.

{\tt Simplify-defun} is one of the transformations of APT (Automated Program
Transformations)~\cite{apt-web}, an ACL2 library of tools to transform programs
and program specifications with a high degree of automation. APT can be used in
program synthesis, to derive provably correct implementations from formal
specifications via sequences of refinement steps carried out via
transformations. APT can also be used in program analysis, to help verify
existing programs, suitably embedded in the ACL2 logic, by raising their level
of abstraction via transformations that are inverses of the ones used in
stepwise program refinement.  In the APT ecosystem, {\tt simplify-defun} is
useful for simplifying and optimizing definitions generated by other APT transformations (e.g.,
transformations that change data representation), as well as to carry out
rewriting transformations via specific sets of rules (e.g., turning unbounded
integer operations into bounded integer operations and vice versa, under
suitable conditions).  It can also be used to chain together and
propagate sequences of transformations (e.g., rewriting a caller by
replacing a callee with a new version, which may in turn be replaced
with an even newer version, and so on).

The idea of using simplification rules to transform programs
is not new~\cite{DBLP:journals/tse/Smith90,DBLP:books/daglib/0096998}.
The contribution of the work described in this paper
is the realization of that idea in ACL2,
which involves techniques that are specific to this prover and environment
and that leverage its existing capabilities and libraries.
This paper could be viewed as a follow-up to an earlier paper on a
related tool~\cite{simplify-defun-2003}.  But that paper focused
largely on usage, while here, we additionally focus both on
interesting applications and on implementation.\footnote{We thank the
  reviewers of the previous paper for suggesting further discussion of
  implementation.  In a sense, we are finally getting around to that!}
Moreover, the new tool was implemented from scratch and improves on
the old tool in several important ways, including the following:

\begin{itemize}

\item The new tool has significantly more options.  Of special note is
  support for patterns that specify which subterms of the
  definition's body to simplify.

\item The new tool has been subjected to a significantly larger
  variety of uses (approximately 300 uses as of mid-January 2017, not
  including artificial tests), illustrating its robustness and
  flexibility.

\item The old tool generated events that were written to a file; calls
  of the new tool are event forms that can be placed in a book.  (The
  old tool pre-dated
  \href{http://www.cs.utexas.edu/users/moore/acl2/manuals/current/manual/index.html?topic=ACL2\_\_\_\_MAKE-EVENT}{\tt\underline{make-event}},
  which is used in the new tool.)

\item The new tool takes advantage of the {\em expander} in community
  book {\tt misc/expander.lisp}, rather than ``rolling its own''
  simplification.  Several recent improvements
  have been made to that book in the course of developing {\tt
    simplify-defun}, which can benefit other users of the expander.

\end{itemize}

{\tt Simplify-defun} may be applied to function symbols that have been
defined using {\tt defun}, possibly with mutual recursion --- perhaps indirectly using a macro, for
example,
\href{http://www.cs.utexas.edu/users/moore/acl2/manuals/current/manual/index.html?topic=ACL2\_\_\_\_DEFUND}{\tt\underline{defund}}
or
\href{http://www.cs.utexas.edu/users/moore/acl2/manuals/current/manual/index.html?topic=ACL2\_\_\_\_DEFINE}{\tt\underline{define}}.
An analogous tool, {\tt simplify-defun-sk}, may be applied to function
symbols defined with
\href{http://www.cs.utexas.edu/users/moore/acl2/manuals/current/manual/index.html?topic=ACL2\_\_\_\_DEFUN-SK}{\tt\underline{defun-sk}},
but we do not discuss it here.

Notice the underlining above.  Throughout this paper, we underline hyperlinks to topics in the online
\href{http://www.cs.utexas.edu/users/moore/acl2/manuals/current/manual/index.html}{\underline{documentation}}~\cite{acl2:doc}
for ACL2 and its books.  We use ACL2 notation freely, abbreviating
with an ellipsis ({\tt ...}) to indicate omitted text and sometimes
modifying whitespace in displayed output.

The rest of this section introduces {\tt simplify-defun} via some very simple
examples that illustrate the essence of the tool.
Section~\ref{sec:examples} presents some examples of
the tool's use in program transformation; that section provides motivation for
some of the features supported by {\tt simplify-defun}.
Section~\ref{sec:options} summarizes options for controlling this tool.  In
Section~\ref{sec:implementation} we discuss how {\tt simplify-defun}
circumvents ACL2's mercurial heuristics and sensitivity to
\href{http://www.cs.utexas.edu/users/moore/acl2/manuals/current/manual/index.html?topic=ACL2\_\_\_\_THEORIES}{\underline{theories}}
so that proofs succeed reliably and automatically.  We conclude in
Section~\ref{sec:conclusion}.

\subsection{Simple Illustrative Examples}

We start with a very simple example that captures much of the essence
of {\tt simplify-defun}.

\begin{verbatim}
(include-book "simplify-defun")
(defun f (x)
  (if (zp x) 0 (+ 1 1 (f (+ -1 x)))))
\end{verbatim}

\noindent Next, we run {\tt simplify-defun} to produce a new
definition and a {\tt defthm} with the formula shown below.  {\bf
  Note}: All {\tt defun} forms generated by {\tt simplify-defun}
contain
\href{http://www.cs.utexas.edu/users/moore/acl2/manuals/current/manual/index.html?topic=ACL2\_\_\_\_DECLARE}{\tt\underline{declare}}
forms, which are generally omitted in this paper; also, whitespace is
liberally edited.

\begin{verbatim}
ACL2 !>(simplify-defun f)
 (DEFUN F{1} (X)
   (IF (ZP X) 0 (+ 2 (F{1} (+ -1 X)))))
ACL2 !>:pf f-becomes-f{1}
(EQUAL (F X) (F{1} X))
ACL2 !>
\end{verbatim}

Several key aspects of a successful {\tt simplify-defun} run are
illustrated above:

\begin{itemize}

\item A new function symbol is defined,
using the \href{http://www.cs.utexas.edu/users/moore/acl2/manuals/current/manual/index.html?topic=ACL2\_\_\_\_NUMBERED-NAMES}{\underline{numbered-names}}
utilities.

\item The body of the new definition is a simplified version of the
  body of the original definition, but with the old function replaced
  by the new in recursive calls.

\item A {\em \becomes} theorem is proved, which states the equivalence of
  the old and new function.

\end{itemize}

This behavior of {\tt simplify-defun} extends naturally to mutual
recursion, in which case a new
\href{http://www.cs.utexas.edu/users/moore/acl2/manuals/current/manual/index.html?topic=ACL2\_\_\_\_MUTUAL-RECURSION}{\tt\underline{mutual-recursion}} event is generated together with \becomes
theorems.  Consider this definition.

\begin{verbatim}
(mutual-recursion
 (defun f1 (x) (if (consp x) (not (f2 (nth 0 x))) t))
 (defun f2 (x) (if (consp x) (f1 (nth 0 x)) t)))
\end{verbatim}

\noindent The result presents no surprises when compared to our first example.

\begin{verbatim}
ACL2 !>(simplify-defun f1)
 (MUTUAL-RECURSION (DEFUN F1{1} (X)
                     (IF (CONSP X) (NOT (F2{1} (CAR X))) T))
                   (DEFUN F2{1} (X)
                     (IF (CONSP X) (F1{1} (CAR X)) T)))
ACL2 !>:pf f1-becomes-f1{1}
(EQUAL (F1 X) (F1{1} X))
ACL2 !>:pf f2-becomes-f2{1}
(EQUAL (F2 X) (F2{1} X))
ACL2 !>
\end{verbatim}

{\tt Simplify-defun} makes some attempt to preserve structure from the
original definitions.  For example, the body of the definition of {\tt f1}
(above) is stored, as usual, as a translated
\href{http://www.cs.utexas.edu/users/moore/acl2/manuals/current/manual/index.html?topic=ACL2\_\_\_\_TERM}{\underline{term}}.
As with most utilities that manipulate ACL2 terms, {\tt
  simplify-defun} operates on translated terms.  So
the new {\tt defun} event form could easily use the transformed body, shown here; notice that {\tt
  T} is quoted.

\begin{verbatim}
ACL2 !>(body 'f1{1} nil (w state))
(IF (CONSP X) (NOT (F2{1} (CAR X))) 'T)
ACL2 !>
\end{verbatim}

\noindent If we naively produced a user-level
(\href{http://www.cs.utexas.edu/users/moore/acl2/manuals/current/manual/index.html?topic=ACL2\_\_\_\_UNTRANSLATE}{\underline{untranslate}}d)
term from that body, the result would look quite different from the
original definition's body.

\begin{verbatim}
ACL2 !>(untranslate (body 'f1{1} nil (w state)) nil (w state))
(OR (NOT (CONSP X)) (NOT (F2{1} (CAR X))))
ACL2 !>
\end{verbatim}

\noindent Therefore, {\tt simplify-defun} uses the
\href{http://www.cs.utexas.edu/users/moore/acl2/manuals/current/manual/index.html?topic=ACL2\_\_\_\_DIRECTED-UNTRANSLATE}{\tt\underline{directed-untranslate}}
utility to untranslate the new (translated) body, heuristically using the
old body (translated and untranslated) as a guide.
This utility was implemented in support of {\tt simplify-defun},
but it is of more general use (e.g., in other APT transformations).

There are many ways to control {\tt simplify-defun} by using
keyword arguments, as described in the next two sections.  Here we
show how to limit simplification to specified subterms.  Consider:

\begin{verbatim}
(defun g (x y)
  (list (+ (car (cons x y)) 3)
        (* (car (cons y y)) 4)
        (* (car (cons x y)) 5)))
\end{verbatim}

\noindent The {\tt :simplify-body} keyword option below specifies
simplification of any occurrence of {\tt (car (cons x y))} that is the first
argument of a call to {\tt *}.  The wrapper {\tt :@} indicates the
simplification site, and the underscore (\verb|_|) matches anything.
Notice that, in the result, only the indicated call is simplified.

\begin{verbatim}
ACL2 !>(simplify-defun g :simplify-body (* (:@ (car (cons x y))) _))
 (DEFUN G{1} (X Y)
   (LIST (+ (CAR (CONS X Y)) 3)
         (* (CAR (CONS Y Y)) 4)
         (* X 5)))
ACL2 !>
\end{verbatim}

\section{Some Applications}
\label{sec:examples}
This section presents some practical examples of use of
\verb|simplify-defun| in program transformation.  They use some keyword
options, which are described in Section~\ref{sec:options} but we hope
are self-explanatory here.  Not shown here are hints generated by
\verb|simplify-defun| to automate proofs, for example by reusing
previous guards, measures, and guard and termination theorems; this is
covered briefly in subsequent sections.

\subsection{Combining a Filter with a Doubly-Recursive Producer}

This example shows how \verb|simplify-defun| is used to apply rewrite rules,
to improve the results of other transformations, and to chain
together previous transformation steps.  The main function {\tt f}
below produces all pairs of items from {\tt x} and {\tt y} and then
filters the result to keep only the ``good'' pairs.

\begin{verbatim}
(defun pair-with-all (item lst) ;; pair ITEM with all elements of LST
  (if (endp lst)
      nil
    (cons (cons item (car lst))
          (pair-with-all item (cdr lst)))))

(defun all-pairs (x y) ;; make all pairs of items from X and Y
  (if (endp x)
      nil
    (append (pair-with-all (car x) y)
            (all-pairs (cdr x) y))))

(defstub good-pair-p (pair) t) ;; just a place holder

(defun keep-good-pairs (pairs)
  (if (endp pairs)
      nil
    (if (good-pair-p (car pairs))
        (cons (car pairs) (keep-good-pairs (cdr pairs)))
      (keep-good-pairs (cdr pairs)))))

(defun f (x y) (keep-good-pairs (all-pairs x y)))
\end{verbatim}

We wish to make {\tt f} more efficient; it should refrain from ever
adding non-good pairs to the result, rather than filtering them out
later.  {\tt F}'s body is {\tt (keep-good-pairs (all-pairs x y))},
which we can improve by ``pushing'' {\tt keep-good-pairs} into the
{\tt if}-branches of {\tt all-pairs}, using APT's {\tt wrap-output}
transformation (not described here). {\tt Wrap-output} produces a
function and a theorem.

\begin{verbatim}
(DEFUN ALL-GOOD-PAIRS (X Y) ; generated by wrap-output
  (IF (ENDP X)
      (KEEP-GOOD-PAIRS NIL)
      (KEEP-GOOD-PAIRS (APPEND (PAIR-WITH-ALL (CAR X) Y)
                               (ALL-PAIRS (CDR X) Y)))))

(DEFTHM RULE1 ; generated by wrap-output
  (EQUAL (KEEP-GOOD-PAIRS (ALL-PAIRS X Y))
         (ALL-GOOD-PAIRS X Y)))
\end{verbatim}

Below, we will apply {\tt rule1} to simplify
{\tt f}.  But first we will further transform {\tt all-good-pairs}. It
can be simplified in three ways.  First, {\tt (keep-good-pairs nil)}
can be evaluated.  Second, we can push the call to {\tt
  keep-good-pairs} over the {\tt append} using this rule.

\begin{verbatim}
(defthm keep-good-pairs-of-append
  (equal (keep-good-pairs (append x y))
         (append (keep-good-pairs x) (keep-good-pairs y))))
\end{verbatim}

Third, note that {\tt all-good-pairs}, despite being a transformed
version of {\tt all-pairs}, is not recursive (it calls the old
function {\tt all-pairs}), but we want it to be recursive.
After {\tt keep-good-pairs} is pushed over the {\tt
  append}, it will be composed with the call of {\tt all-pairs}, which
is the exact pattern that {\tt rule1} can rewrite to a call to {\tt
  all-good-pairs}.  {\tt Simplify-defun} applies these
simplifications.

\begin{verbatim}
ACL2 !>(simplify-defun all-good-pairs)
 (DEFUN ALL-GOOD-PAIRS{1} (X Y)
   (IF (ENDP X)
       NIL
       (APPEND (KEEP-GOOD-PAIRS (PAIR-WITH-ALL (CAR X) Y))
               (ALL-GOOD-PAIRS{1} (CDR X) Y))))
ACL2 !>
\end{verbatim}

Note that the new function is recursive.  This is because
{\tt rule1} introduced a call to {\tt all-good-pairs}, which
{\tt simplify-defun} then renamed to \verb|all-good-pairs{1}| (it always renames
recursive calls). We have made some progress pushing {\tt keep-good-pairs} closer
to where the pairs are created. Now, \verb|all-good-pairs{1}| can be further
simplified.  Observe that its body contains composed calls of
{\tt keep-good-pairs} and {\tt pair-with-all}.  We can optimize this term using APT's
{\tt producer-consumer} transformation (not described here) to combine the
creation of the pairs with the filtering of good pairs.  As usual, a new
function and a theorem are produced.

\begin{verbatim}
(DEFUN PAIR-WITH-ALL-AND-FILTER (ITEM LST) ; generated by producer-consumer
  (IF (ENDP LST)
      NIL
      (IF (GOOD-PAIR-P (CONS ITEM (CAR LST)))
          (CONS (CONS ITEM (CAR LST))
                (PAIR-WITH-ALL-AND-FILTER ITEM (CDR LST)))
          (PAIR-WITH-ALL-AND-FILTER ITEM (CDR LST)))))

(DEFTHM RULE2 ; generated by producer-consumer
  (EQUAL (KEEP-GOOD-PAIRS (PAIR-WITH-ALL ITEM LST))
         (PAIR-WITH-ALL-AND-FILTER ITEM LST)))
\end{verbatim}

{\tt Pair-with-all-and-filter} immediately discards non-good pairs, saving work compared to filtering them out later.
Now {\tt simplify-defun} can change \verb|all-good-pairs{1}| by
applying {\tt rule2}.

\begin{verbatim}
ACL2 !>(simplify-defun all-good-pairs{1})
 (DEFUN ALL-GOOD-PAIRS{2} (X Y)
   (IF (ENDP X)
       NIL
       (APPEND (PAIR-WITH-ALL-AND-FILTER (CAR X) Y)
               (ALL-GOOD-PAIRS{2} (CDR X) Y))))
ACL2 !>
\end{verbatim}

Finally, we apply {\tt simplify-defun} to transform {\tt f} by applying all of the
preceding rewrites in succession, introducing {\tt all-good-pairs}, which is in
turn replaced with \verb|all-good-pairs{1}| and then \verb|all-good-pairs{2}|.

\begin{verbatim}
ACL2 !>(simplify-defun f :new-name f-fast)
 (DEFUN F-FAST (X Y)
   (ALL-GOOD-PAIRS{2} X Y))
ACL2 !>:pf f-becomes-f-fast
(EQUAL (F X Y) (F-FAST X Y))
\end{verbatim}

This builds a fast version of {\tt f} and a theorem proving it equal to {\tt f}.

\subsection{Converting between Unbounded and Bounded Integer Operations}

Popular programming languages like C and Java
typically use bounded integer types and operations,
while requirements specifications
typically use unbounded integer types and operations.
Thus, synthesizing a C or Java program from a specification,
or proving that a C or Java program complies with a specification,
often involves showing that unbounded and bounded integers are ``equivalent''
under the preconditions stated by the specification.

Consider this Java implementation
of Bresenham's line drawing algorithm~\cite{DBLP:journals/ibmsj/Bresenham65}
for the first octant.%
\footnote{This algorithm computes a best-fit discrete line
  using only integer operations.
  Understanding the algorithm is not necessary
  for the purpose of this {\tt simplify-defun} example.}

\begin{verbatim}
// draw a line from (0, 0) to (a, b), where 0 <= b <= a <= 1,000,000:
static void drawLine(int a, int b) {
    int x = 0, y = 0, d = 2 * b - a;
    while (x <= a) {
        drawPoint(x, y); // details unimportant
        x++;
        if (d >= 0) { y++; d += 2 * (b - a); }
        else { d += 2 * b; }
    }
}
\end{verbatim}

Assuming the screen width and height are less than 1,000,000 pixels,
none of the two's complement 32-bit integer operations in the Java method
 wrap around. So they
 could be replaced with corresponding unbounded integer operations,
as shown below.
This replacement raises the level of abstraction
and helps verify the functional correctness of the method.

The Java code above can be represented as shown below in ACL2
(see the paper's supporting materials for full details), where:

\begin{itemize}
\item
  \verb|Int32p| recognizes
  some representation of Java's two's complement 32-bit integers,
  whose details are unimportant.
\item
  \verb|Int32| converts an ACL2 integer in $[-2^{31},2^{31})$
  (i.e., an \verb|x| such that \verb|(signed-byte-p 32 x)| holds)
  to the corresponding representation in \verb|int32p|.
\item
  \verb|Int| converts a representation in \verb|int32p|
  to the corresponding ACL2 integer in $[-2^{31},2^{31})$.
\item
  \verb|Add32|, \verb|sub32|, and \verb|mul32|
  represent Java's two's complement 32-bit
  addition, subtraction, and multiplication operations.
\item
  \verb|Lte32| and \verb|gte32|
  represent Java's two's complement 32-bit
  less-than-or-equal-to and greater-than-or-equal-to operations.
\item
  \verb|Drawline-loop| represents the loop as a state transformer,
  whose state consists of
  \verb|a|, \verb|b|, \verb|x|, \verb|y|, \verb|d|, and the screen.
  This function is ``defined'' only where the loop invariant holds.
  The guard verification of this function implies
  the preservation of the loop invariant.
\item
  \verb|Drawline| represents the method as a function that maps
  \verb|a|, \verb|b|, and the current screen to an updated screen.
  This function is ``defined''only where the precondition holds.
  The guard verification of this function implies
  the establishment of the loop invariant.
\end{itemize}
The Axe tool~\cite{smith-axe} can automatically generate
a representation similar to this one from Java (byte)code,
with some additional input from the user (e.g., part of the loop invariant).

\begin{verbatim}
(defun drawpoint (x y screen)
  (declare (xargs :guard (and (int32p x) (int32p y))))
  ...) ; returns updated screen, details unimportant

(defun precond (a b) ; precondition of the method
  (declare (xargs :guard t))
  (and (int32p a) ; Java type of a
       (int32p b) ; Java type of b
       (<= 0 (int b))
       (<= (int b) (int a))
       (<= (int a) 1000000)))

(defun invar (a b x y d) ; loop invariant of the method
  (declare (xargs :guard t))
  (and (precond a b)
       (int32p x) ; Java type of x
       (int32p y) ; Java type of y
       (int32p d) ; Java type of d
       ...)) ; conditions on x, y, and d, details unimportant

(defun drawline-loop (a b x y d screen) ; loop of the method
  (declare (xargs :guard (invar a b x y d)
                  ...)) ; measure and (guard) hints, details unimportant
  (if (invar a b x y d)
      (if (not (lte32 x a))
          screen ; exit loop
        (drawline-loop a b
                       (add32 x (int32 1))
                       (if (gte32 d (int32 0))
                           (add32 y (int32 1))
                         y)
                       (if (gte32 d (int32 0))
                           (add32 d (mul32 (int32 2) (sub32 b a)))
                         (add32 d (mul32 (int32 2) b)))
                       (drawpoint x y screen)))
    :undefined))

(defun drawline (a b screen) ; method
  (declare (xargs :guard (precond a b)
                  ...)) ; guard hints, details unimportant
  (if (precond a b)
      (drawline-loop a b
                     (int32 0) ; x
                     (int32 0) ; y
                     (sub32 (mul32 (int32 2) b) a) ; d
                     screen)
    :undefined))
\end{verbatim}

The following rewrite rules are disabled
because their right-hand sides are not generally ``simpler'' or ``better''
than their left-hand sides.
But when passed to the \verb|:enable| option of \verb|simplify-defun|,
which instructs the tool to use these rules in the expander,
they systematically replace bounded integer operations
with their unbounded counterparts.

\begin{verbatim}
(defthmd add32-to-+  (equal (add32 x y) (int32 (+ (int x) (int y)))))
(defthmd sub32-to--  (equal (sub32 x y) (int32 (- (int x) (int y)))))
(defthmd mul32-to--  (equal (mul32 x y) (int32 (* (int x) (int y)))))
(defthmd lte32-to-<= (equal (lte32 x y) (<= (int x) (int y))))
(defthmd gte32-to-<= (equal (gte32 x y) (>= (int x) (int y))))
\end{verbatim}

Since these rewrite rules are unconditional,
the replacement always occurs,
but subterms of the form \verb|(int (int32 ...))| are generated.
For instance,
the term \verb|(add32 d (mul32 (int32 2) b))| above becomes
\verb|(int32 (+ (int d) (int (int32 (* (int (int32 2)) (int b))))))|.
These \verb|(int (int32 ...))| terms can be simplified
via the following conditional rewrite rule
(which the expander in \verb|simplify-defun| uses by default,
since it is an enabled rule).
Relieving the hypotheses of this rewrite rule's applicable instances
amounts to showing that each bounded integer operation does not wrap around
in the expressions under consideration.

\begin{verbatim}
(defthm int-of-int32
  (implies (signed-byte-p 32 x)
           (equal (int (int32 x)) x)))
\end{verbatim}

Applying \verb|simplify-defun| to \verb|drawline-loop| and \verb|drawline|
yields the desired results.
In this case, the \verb|int-of-int32| hypotheses are automatically relieved.
These uses of \verb|simplify-defun| show a practical use of the pattern feature:
\verb|invar| and \verb|precond| must be enabled
to relieve the \verb|int-of-int32| hypotheses,
but we want the generated function to keep them unopened;
so we use a pattern that limits the simplification
to the true branches of the \verb|if|s.
The \becomes theorem generated by the first \verb|simplify-defun|
is used by the second to have \verb|drawline{1}|
call \verb|drawline-loop{1}| instead of \verb|drawline-loop|;
this is another example of propagating transformations.

\begin{verbatim}
ACL2 !> (simplify-defun drawline-loop
                        :simplify-body (if _ @ _)
                        :enable (add32-to-+ ... gte32-to->=))
 (DEFUN DRAWLINE-LOOP{1} (A B X Y D SCREEN)
   (DECLARE ...)
   (IF (INVAR A B X Y D)
       (IF (NOT (< (INT A) (INT X)))
             (DRAWLINE-LOOP{1} A B
                             (INT32 (+ 1 (INT X)))
                             (IF (< (INT D) 0)
                                  Y
                               (INT32 (+ 1 (INT Y))))
                             (IF (< (INT D) 0)
                                 (INT32 (+ (INT D) (* 2 (INT B))))
                               (INT32 (+ (INT D)
                                         (- (* 2 (INT A)))
                                         (* 2 (INT B)))))
                             (DRAWPOINT X Y SCREEN))
     SCREEN)
    :UNDEFINED))
ACL2 !> (simplify-defun drawline
                        :simplify-body (if _ @ _)
                        :enable (add32-to-+ ... gte32-to->=))
 (DEFUN DRAWLINE{1} (A B SCREEN)
   (DECLARE ...)
   (IF (PRECOND A B)
       (DRAWLINE-LOOP{1} A B
                         (INT32 0)
                         (INT32 0)
                         (INT32 (+ (- (INT A)) (* 2 (INT B))))
                         SCREEN)
     :UNDEFINED))
ACL2 !>
\end{verbatim}

The resulting expressions have
the \verb|int| conversion at the variable leaves
and the \verb|int32| conversion at the roots.
APT's isomorphic data transformation (not discussed here)
can eliminate them by changing the data representation
of the functions' arguments,
generating functions that no longer deal with bounded integers.
The resulting functions are more easily proved
to satisfy the high-level functional specification of the algorithm,
namely that it produces a best-fit discrete line
(this proof is not discussed here).

The bounded-to-unbounded operation rewriting technique shown here,
followed by the isomorphic data transformation mentioned above,
should have general applicability.
Proving that bounded integer operations do not wrap around
may be arbitrarily hard:
when the \verb|int-of-int32| hypotheses cannot be relieved automatically,
the user may have to prove lemmas to help \verb|simplify-defun|.
When a Java computation is supposed to wrap around
(e.g., when calculating a hash),
the specification must explicitly say that,
and slightly different rewrite rules may be needed.
When synthesizing code (as opposed to analyzing code as in this example),
it should be possible to use a similar technique
with rewrite rules that turn
unbounded integer operations into their bounded counterparts,
``inverses'' of the rules \verb|add32-to-+|, \verb|sub32-to--|, etc.

\section{Options}
\label{sec:options}
This section very briefly summarizes the keyword arguments of {\tt
  simplify-defun}.  Here we assume that the given function's
  definition is not mutually recursive; for that case and other
  details, see the
  \href{http://www.cs.utexas.edu/users/moore/acl2/manuals/current/manual/index.html?topic=ACL2\_\_\_\_XDOC}{\underline{XDOC}}
  topic for simplify-defun, provided by the supporting materials.

{\bf Assumptions.}  A list of assumptions under which the body is
simplified can be specified by the {\tt :assumptions} keyword.
Or, keyword {\tt :hyp-fn} can specify the assumptions using
a function symbol.

{\bf Controlling the Result.}  By default, the new function symbol is
\href{http://www.cs.utexas.edu/users/moore/acl2/manuals/current/manual/index.html?topic=ACL2\_\_\_\_DISABLE}{\underline{disable}}d
if and only if the input function symbol is disabled.  However, that
default can be overridden with keyword {\tt :function-disabled}.
Similarly, the measure, guard verification, and non-executability
status come from the old function symbol but can be overridden by
keywords {\tt :measure}, {\tt :verify-guards}, and {\tt
  :non-executable}, respectively.  The \becomes theorem is
\href{http://www.cs.utexas.edu/users/moore/acl2/manuals/current/manual/index.html?topic=ACL2\_\_\_\_ENABLE}{\underline{enable}}d
by default, but this can be overridden by keyword {\tt
  :theorem-disabled}.  Keywords can also specify the names for the new
function symbol ({\tt :new-name}) and \becomes theorem ({\tt
  :theorem-name}).  The new function body is produced, by default, using the
\href{http://www.cs.utexas.edu/users/moore/acl2/manuals/current/manual/index.html?topic=ACL2\_\_\_\_DIRECTED-UNTRANSLATE}{\tt\underline{directed-untranslate}}
utility (see Section~\ref{sec:intro}); but keyword {\tt
  :untranslate} can specify to use the ordinary untranslate operation
or even to leave the new body in translated form.  Finally, by default
the new function produces results equal to the old; however, an
equivalence relation between the old and new results can be specified
with keyword {\tt :equiv}, which is used in the statement of the
\becomes theorem.

{\bf Specifying Theories.}  The {\tt :theory} keyword specifies the
theory to be used when simplifying the definition; alternatively, {\tt
  :disable} and {\tt :enable} keywords can be used for this purpose.
Keyword {\tt :expand} can be used with any (or none) of these to
specify terms to expand, as with ordinary {\tt :expand} hints for the
prover.  Similarly, there are keywords {\tt :assumption-theory}, {\tt
  :assumption-disable}, and {\tt :assumption-enable} for controlling
the theory used when proving that assumptions are preserved by
recursive calls (an issue discussed in Section~\ref{sec:assumptions}).
There are also keywords {\tt :measure-hints} and {\tt :guard-hints}
with the obvious meanings.

{\bf Specifying Simplification.}  By default, {\tt simplify-defun}
attempts to simplify the body of the given function symbol, but not
its guard or measure.  Keywords {\tt :simplify-body}, {\tt
  :simplify-guard}, and {\tt :simplify-measure} can override that
default behavior.  By default, {\tt simplify-defun} fails if
it attempts to simplify the body but fails to do so, though there is
no such requirement for the guard or measure; keyword {\tt
  :must-simplify} can override those defaults.

{\bf Output Options.}  The keyword {\tt :show-only} causes {\tt
  simplify-defun} not to change the
\href{http://www.cs.utexas.edu/users/moore/acl2/manuals/current/manual/?topic=ACL2____WORLD}{\underline{world}}, but instead to show how it
expands into primitive events (see Section~\ref{sec:implementation}).
If {\tt :show-only} is {\tt nil} (the default), then by default, the
new definition is printed when {\tt simplify-defun} is successful;
keyword {\tt :print-def} can suppress that printing.  Finally, a {\tt
  :verbose} option can provide extra output.

\section{Implementation}
\label{sec:implementation}
{\tt Simplify-defun} is designed to apply the {\em expander},
specifically, function {\tt tool2-fn} in community book {\tt
  misc/expander.lisp}, which provides an interface to the
ACL2 rewriter in a context based on the use of forward-chaining
and linear arithmetic.  The goal is to simplify the definition as
specified and to arrange that all resulting proofs succeed fully
automatically and quickly.

This section discusses how {\tt simplify-defun} achieves this goal,
with a few (probably infrequent) exceptions in the case of proofs for
guards, termination, or (discussed below) that assumptions are
preserved by recursive calls.  First, we explain the full form
generated by a call of {\tt simplify-defun}, which we call its {\em
  expansion}; this form carefully orchestrates the proofs.  Then we
dive deeper by exploring the proof of the \becomes theorem, focusing on
the use of functional instantiation.  Next we see how the expansion is
modified when assumptions are used.  Finally we provide a
few brief implementation notes.

One can experiment using the supporting materials: see the book {\tt
  simplify-defun-tests.lisp}; or simply include the book {\tt
  simplify-defun}, define a function {\tt f}, and then evaluate {\tt
  (simplify-defun f :show-only t)}, perhaps adding options, to see the
expansion.

The implementation takes advantage of many features offered by ACL2
for system building (beyond its prover engine), including for example
\href{http://www.cs.utexas.edu/users/moore/acl2/manuals/current/manual/index.html?topic=ACL2\_\_\_\_ENCAPSULATE}{\tt\underline{encapsulate}},
\href{http://www.cs.utexas.edu/users/moore/acl2/manuals/current/manual/index.html?topic=ACL2\_\_\_\_MAKE-EVENT}{\tt\underline{make-event}},
\href{http://www.cs.utexas.edu/users/moore/acl2/manuals/current/manual/index.html?topic=ACL2\_\_\_\_WITH-OUTPUT}{\tt\underline{with-output}}.
We say a bit more about this at the end of the section.

\subsection{The {\tt Simplify-defun} Expansion}

We illustrate the expansion generated by {\tt simplify-defun} using
the following definition, which is the first example of
Section~\ref{sec:intro} but with a guard added.

\begin{verbatim}
(defun f (x)
 (declare (xargs :guard (natp x)))
 (if (zp x) 0 (+ 1 1 (f (+ -1 x)))))
\end{verbatim}

\noindent The expansion is an
\href{http://www.cs.utexas.edu/users/moore/acl2/manuals/current/manual/index.html?topic=ACL2\_\_\_\_ENCAPSULATE}{\tt\underline{encapsulate}}
event.  We can see the expansion by evaluating the form {\tt
  (simplify\-/defun f :show-only t)}, which includes the indicated
four sections to be discussed below.

\begin{Verbatim}[commandchars=\\\{\},fontsize=\small]
(encapsulate nil
 {\em{prelude}}
 {\em{new defun form}}
 (local (progn {\em{local events}}))
 {\em{\becomes theorem}})
\end{Verbatim}

\subsubsection{Prelude}
\label{sec:prelude}

The prelude is mostly independent of {\tt f}, but the name of {\tt f}
is supplied to {\tt install-not-normalized}.

\begin{Verbatim}[commandchars=\\\{\},fontsize=\small]
(SET-INHIBIT-WARNINGS "theory")
(SET-IGNORE-OK T)
(SET-IRRELEVANT-FORMALS-OK T)
(LOCAL (INSTALL-NOT-NORMALIZED F))
(LOCAL (SET-DEFAULT-HINTS NIL))
(LOCAL (SET-OVERRIDE-HINTS NIL))
\end{Verbatim}

The first form above avoids warnings due to the use of small theories for
directing the prover.  The next two forms allow simplification to make
some formals unused or irrelevant in the new definition.  The use of
\href{http://www.cs.utexas.edu/users/moore/acl2/manuals/current/manual/index.html?topic=ACL2\_\_\_\_INSTALL-NOT-NORMALIZED}{\tt\underline{install-not-normalized}}
is a bit subtle perhaps, but not complicated: by default, ACL2 stores
a simplified, or
{\em\href{http://www.cs.utexas.edu/users/moore/acl2/manuals/current/manual/index.html?topic=ACL2\_\_\_\_NORMALIZE}{\underline{normalize}}d},
body for a function symbol, but {\tt simplify-defun} is intended to generate
a definition based on the body $b$ of the old definition as it was submitted, not on
the normalization of $b$.  Using {\tt install-not-normalized}
arranges that {\tt :expand} hints for {\tt f} will use the
unnormalized body.  This supports the proofs generated by {\tt
  simplify-defun}, by supporting reasoning about the unnormalized
bodies of both the old and new functions.
Finally, the last two forms
guarantee that the global environment will not sabotage the proof.

Space does not permit discussion of the handling of
\href{http://www.cs.utexas.edu/users/moore/acl2/manuals/current/manual/index.html?topic=ACL2\_\_\_\_MUTUAL-RECURSION}{\tt\underline{mutual-recursion}}.
We mention here only that the form {\tt (SET-BOGUS-MUTUAL-RECURSION-OK
  T)} is then added to the prelude, in case some of the (mutual)
recursion disappears with simplification.

\subsubsection{New Defun Form}

The following new definition has the same simplified body as for the
corresponding example in Section~\ref{sec:intro}.  As before, the new
body is produced by running the expander on the unnormalized body of
{\tt f}.

\begin{verbatim}
(DEFUN F{1} (X)
  (DECLARE (XARGS :NORMALIZE NIL
                  :GUARD (NATP X)
                  :MEASURE (ACL2-COUNT X)
                  :VERIFY-GUARDS T
                  :GUARD-HINTS (("Goal" :USE (:GUARD-THEOREM F)) ...)
                  :HINTS (("Goal" :USE (:TERMINATION-THEOREM F)) ...)))
  (IF (ZP X) 0 (+ 2 (F{1} (+ -1 X)))))
\end{verbatim}

The
\href{http://www.cs.utexas.edu/users/moore/acl2/manuals/current/manual/index.html?topic=ACL2\_\_\_\_GUARD}{\underline{guard}}
and
\href{http://www.cs.utexas.edu/users/moore/acl2/manuals/current/manual/index.html?topic=ACL2\_\_\_\_MEASURE}{\underline{measure}}
are (by default) inherited from {\tt f}.  Since {\tt f} is
guard-verified, then by default, so is the new function.
The uses of
\href{http://www.cs.utexas.edu/users/moore/acl2/manuals/current/manual/index.html?topic=ACL2\_\_\_\_GUARD-THEOREM}{\tt{:{\underline{GUARD-THEOREM}}}}
and
\href{http://www.cs.utexas.edu/users/moore/acl2/manuals/current/manual/index.html?topic=ACL2\_\_\_\_TERMINATION-THEOREM}{\tt{:{\underline{TERMINATION-THEOREM}}}}
are designed to make the guard and termination proofs automatic and
fast in most cases, and were implemented in support of the APT project.

\subsubsection{Local Events}
\label{sec:local-events}

The local events are how {\tt simplify-defun} carefully arranges for
proofs to succeed automatically and fast, in most cases.  The first
local event defines a
\href{http://www.cs.utexas.edu/users/moore/acl2/manuals/current/manual/index.html?topic=ACL2\_\_\_\_THEORY}{\underline{theory}}
to be used when proving equality of the body of {\tt f} with a
simplified version.  It is the union of the
\href{http://www.cs.utexas.edu/users/moore/acl2/manuals/current/manual/index.html?topic=ACL2\_\_\_\_RUNES}{\underline{runes}}
reported by the expander when simplifying the body, with the set of
all {\tt
  :\href{http://www.cs.utexas.edu/users/moore/acl2/manuals/current/manual/index.html?topic=ACL2\_\_\_\_CONGRUENCE}{\underline{congruence}}}
and {\tt
  :\href{http://www.cs.utexas.edu/users/moore/acl2/manuals/current/manual/index.html?topic=ACL2\_\_\_\_EQUIVALENCE}{\underline{equivalence}}}
runes, since these are not tracked by the ACL2 rewriter.\footnote{We
  originally generated a form {\tt (defconst *f-runes* ...)} that
  listed all runes from the {\tt union-equal} call; but the congruence
  theory made that list long and distracting when viewing the
  expansion with {\tt :show-only t}.  Note that the congruence theory
  includes {\tt :equivalence} runes, which after all represent
  congruence rules for diving into calls of equivalence relations.}

\begin{verbatim}
(MAKE-EVENT (LET ((THY (UNION-EQUAL '((:REWRITE FOLD-CONSTS-IN-+)
                                      (:EXECUTABLE-COUNTERPART BINARY-+)
                                      (:DEFINITION SYNP))
                                    (CONGRUENCE-THEORY (W STATE)))))
                 (LIST 'DEFCONST '*F-RUNES* (LIST 'QUOTE THY))))
\end{verbatim}

\noindent The second local event proves the equality of the body with
its simplified version.  Notice that the latter is still in terms of
{\tt f}; the new function symbol, \verb|f{1}|, will be introduced
later.  The
\href{http://www.cs.utexas.edu/users/moore/acl2/manuals/current/manual/index.html?topic=ACL2\_\_\_\_PROOF-BUILDER}{\underline{proof-builder}}
{\tt :instructions} are carefully generated to guarantee that the
proof succeeds; in this simple example, they first put the
proof-builder in the smallest theory that should suffice (for
efficiency), then simplify the old body (the first argument of the
{\tt equal} call), and then prove the resulting equality (which at
that point should be the equality of two identical terms).

\begin{verbatim}
(DEFTHM F-BEFORE-VS-AFTER-0
  (EQUAL (IF (ZP X) 0 (+ 1 1 (F (+ -1 X))))
         (IF (ZP X) 0 (+ 2 (F (+ -1 X)))))
  :INSTRUCTIONS ...
  :RULE-CLASSES NIL)
\end{verbatim}

\noindent The third local event is as follows.

\begin{verbatim}
(COPY-DEF F{1}
          :HYPS-FN NIL
          :HYPS-PRESERVED-THM-NAMES NIL
          :EQUIV EQUAL)
\end{verbatim}

\noindent This macro call introduces a constrained function symbol,
\verb|f{1}-copy|, whose constraint results from the definitional axiom
for \verb|f{1}| by replacing \verb|f{1}| with \verb|f{1}-copy|.  It
also proves the two functions equivalent using a trivial induction in
a tiny theory, to make the proof reliable and fast.

\begin{verbatim}
(DEFTHM F{1}-COPY-DEF
  (EQUAL (F{1}-COPY X)
         (IF (ZP X)
             '0
             (BINARY-+ '2 (F{1}-COPY (BINARY-+ '-1 X)))))
  :HINTS ...
  :RULE-CLASSES ((:DEFINITION :INSTALL-BODY T
                              :CLIQUE (F{1}-COPY)
                              :CONTROLLER-ALIST ((F{1}-COPY T)))))

(LOCAL (IN-THEORY '((:INDUCTION F{1}) F{1}-COPY-DEF (:DEFINITION F{1}))))

(DEFTHM F{1}-IS-F{1}-COPY
  (EQUAL (F{1} X) (F{1}-COPY X))
  :HINTS (("Goal" :INDUCT (F{1} X)))
  :RULE-CLASSES NIL)
\end{verbatim}

\noindent The last local event is the lemma for proving the \becomes
theorem.  In Section~\ref{sec:fn-inst} below we explore this use of
functional instantiation.

\begin{verbatim}
(DEFTHM F-BECOMES-F{1}-LEMMA
  (EQUAL (F{1} X) (F X))
  :HINTS (("Goal"
           :BY (:FUNCTIONAL-INSTANCE F{1}-IS-F{1}-COPY (F{1}-COPY F))
           :IN-THEORY (UNION-THEORIES (CONGRUENCE-THEORY WORLD)
                                      (THEORY 'MINIMAL-THEORY)))
          '(:USE (F-BEFORE-VS-AFTER-0 F$NOT-NORMALIZED))))
\end{verbatim}

\subsubsection{`Becomes' Theorem} 
\label{sec:becomes}

The \becomes theorem in this example states the same theorem as
its lemma above (though we will see in
Section~\ref{sec:assumptions} that this is not always be the case).
The {\tt :in-theory} hint serves to keep
the ACL2 rewriter from bogging down during the proof.

\begin{verbatim}
(DEFTHM F-BECOMES-F{1}
  (EQUAL (F X) (F{1} X))
  :HINTS (("Goal" :USE F-BECOMES-F{1}-LEMMA :IN-THEORY NIL)))
\end{verbatim}

\subsection{Proving the `Becomes' Theorem} 
\label{sec:fn-inst}

Next we see how functional instantiation is used in proving the
\becomes theorem above, or more precisely, its local lemma.  Recall that
above, a
{\tt:\href{http://www.cs.utexas.edu/users/moore/acl2/manuals/current/manual/index.html?topic=ACL2\_\_\_\_BY}{\underline{by}}}
hint is used that replaces \verb|f{1}-copy| by \verb|f| in the lemma
\verb|f{1}-is-f{1}-copy|, {\tt (EQUAL (F\{1\} X) (F\{1\}-COPY X))}, to
prove: {\tt (EQUAL (F X) (F\{1\} X))}.  That substitution works
perfectly (modulo commuting the equality, which the {\tt :by} hint
tolerates), but it requires proving the following property, which
states that \verb|f| satisfies the constraint for \verb|f{1}-copy|.

\begin{verbatim}
(EQUAL (f X)
       (IF (ZP X)
           '0
           (BINARY-+ '2 (f (BINARY-+ '-1 X)))))
\end{verbatim}

\noindent But the right-hand side is exactly what was produced by
applying the expander to the body of {\tt f}.  If we look at the {\tt
  :hints} in \verb|f-becomes-f{1}-lemma| above, we see that after
using functional instantiation (with the {\tt :by} hint), a computed
hint completes the proof by using two facts: {\tt f-before-vs-after-0}
(the second local lemma above), which equates the two bodies; and
\verb|f$not-normalized|, which equates {\tt f} with its unnormalized
(i.e., user-supplied) body.  The latter was created in the prelude
(Section~\ref{sec:prelude})
by the form {\tt (install-not-normalized f)}.
Notice that no proof by induction was
performed for the \becomes theorem (or its local lemma), even though it is
inherently an inductive fact stating the equivalence of
recursive functions \verb|f| and \verb|f{1}|.  We are essentially
taking advantage of the trivial induction already performed in the
proof of the lemma being functionally instantiated,
\verb|f{1}-is-f{1}-copy|.

\subsection{Assumptions}
\label{sec:assumptions}

The following trivial example shows how assumptions change the {\tt
  simplify-defun} expansion.

\begin{verbatim}
(defun foo (x)
    (declare (xargs :guard (true-listp x)))
    (if (consp x)
        (foo (cdr x))
      x))
\end{verbatim}

\noindent Under the {\tt :assumption} of the guard, {\tt (true-listp x)},
the variable {\tt x} in the body is simplified to the constant {\tt
  nil}, using its context {\tt (not (consp x))}.

\begin{Verbatim}[commandchars=\\\{\},fontsize=\small]
ACL2 !>(simplify-defun foo :assumptions :guard :show-only t)
(ENCAPSULATE NIL
 {\em{prelude (as before)}}
 {\em{; new defun form:}}
 (DEFUN FOO\{1\} (X)
   (IF (CONSP X) (FOO\{1\} (CDR X)) NIL))
 (LOCAL (PROGN {\em{local events}})) {\em ; discussed below}
 {\em{; \becomes theorem:}}
 (DEFTHM FOO-BECOMES-FOO\{1\}
   (IMPLIES (TRUE-LISTP X)
            (EQUAL (FOO X) (FOO\{1\} X)))
   :HINTS ...)
ACL2 !>
\end{Verbatim}

\noindent This time, the \becomes theorem has a hypothesis provided
by the {\tt :assumptions}.

We next discuss some differences in the local events generated when there
are assumptions.  A new local event defines a function for the
assumptions, so that when the assumptions are complicated, disabling
that function can hide complexity from the rewriter.

\begin{verbatim}
(DEFUN FOO-HYPS (X) (TRUE-LISTP X))
\end{verbatim}

In order to prove equivalence of the old and new functions, which
involves a proof by induction at some point, it is necessary to reason
that the assumptions are preserved by recursive calls.  The following
local lemma is generated for that purpose.

\begin{verbatim}
(DEFTHM FOO-HYPS-PRESERVED-FOR-FOO
  (IMPLIES (AND (FOO-HYPS X) (CONSP X))
           (FOO-HYPS (CDR X)))
  :HINTS (("Goal" :IN-THEORY (DISABLE* FOO (:E FOO) (:T FOO))
                  :EXPAND ((:FREE (X) (FOO-HYPS X)))
                  :USE (:GUARD-THEOREM FOO)))
  :RULE-CLASSES NIL)
\end{verbatim}

\noindent In many cases the proof of such a lemma will be automatic.
Otherwise, one can first define {\tt foo-hyps} and prove this
lemma before running {\tt simplify-defun}.

There are some tricky wrinkles in the presence of assumptions that we
do not discuss here, in particular how they can affect the use of {\tt
  copy-def} (discussed above in Section~\ref{sec:local-events}).  Some relevant discussion may be found in the
``Essay on the Implementation of Simplify-defun'' in the file {\tt
  simplify-defun.lisp}.

\subsection{Implementation Notes}

The discussion above explains the expansion from a call of {\tt
  simplify-defun}.  Here we discuss at a high level how such forms are
generated.  Consider the first example from the introduction, below,
and let us see its single-step macroexpansion.

\begin{verbatim}
(defun f (x)
  (if (zp x) 0 (+ 1 1 (f (+ -1 x)))))

ACL2 !>:trans1 (simplify-defun f)
 (WITH-OUTPUT :GAG-MODE NIL :OFF :ALL :ON ERROR
   (MAKE-EVENT (SIMPLIFY-DEFUN-FN 'F 'NIL 'NIL ':NONE ... STATE)))
ACL2 !>
\end{verbatim}

\noindent This use of
\href{http://www.cs.utexas.edu/users/moore/acl2/manuals/current/manual/index.html?topic=ACL2\_\_\_\_WITH-OUTPUT}{\tt\underline{with-output}}
prevents output unless there is an error.  The
\href{http://www.cs.utexas.edu/users/moore/acl2/manuals/current/manual/index.html?topic=ACL2\_\_\_\_MAKE-EVENT}{\tt\underline{make-event}}
call instructs {\tt simplify-defun-fn} to produce an event
of the form {\tt (progn $E$ $A$ (value-triple '$D$))},
where $E$ is the expansion, $A$ is a
\href{http://www.cs.utexas.edu/users/moore/acl2/manuals/current/manual/index.html?topic=ACL2\_\_\_\_TABLE}{\tt\underline{table}}
event provided to support redundancy for {\tt simplify-defun}, and $D$
is the new definition (which is thus printed to the terminal).  {\tt
  Simplify-defun-fn} first calls the expander to produce a simplified
body, which it then uses to create $D$ and $E$.  For details see {\tt
  simplify-defun.lisp} in the supporting materials, which we hope is
accessible to those having a little familiarity with ACL2 system
programming (see for example
\href{http://www.cs.utexas.edu/users/moore/acl2/manuals/current/manual/index.html?topic=ACL2\_\_\_\_SYSTEM-UTILITIES}{\underline{system-utilities}}
and
\href{http://www.cs.utexas.edu/users/moore/acl2/manuals/current/manual/index.html?topic=ACL2\_\_\_\_PROGRAMMING-WITH-STATE}{\underline{programming-with-state}}).
These details help proofs to succeed efficiently, in particular by
generating suitable hints, including small theories.  Another detail
is that if the old definition specifies
\href{http://www.cs.utexas.edu/users/moore/acl2/manuals/current/manual/index.html?topic=ACL2\_\_\_\_RULER-EXTENDERS}{\underline{ruler-extenders}}
other than the default, then these are carried over to the new
definition.

\section{Conclusion}
\label{sec:conclusion}
The use of simplification, particularly rewrite rules,
is an old and important idea in program transformation.
\verb|Simplify-defun| realizes this idea in ACL2,
by leveraging the prover's existing
proof procedures, libraries, and environment.
It is one of the transformations of the APT tool suite
for transforming programs and program specifications,
useful for both synthesis and verification.
While \texttt{simplify-defun} is appropriate for equivalence-preserving refinements,
other APT transformations are appropriate for other kinds of refinement.
For instance, specifications that allow more than one implementation
(see~\cite[Section 2]{soft-2015} for an example)
can be refined via APT's narrowing transformation (not discussed here).

We have used \verb|simplify-defun| quite extensively in program derivations,
demonstrating its robustness and utility.
This paper describes not only the general usage of \verb|simplify-defun|,
but also ACL2-specific techniques used to implement it.
It also shows some non-trivial examples that illustrate the tool's utility.

The tool continues to evolve.  Developments since the drafting
of this paper include: delaying guard verification; and expanding
all {\tt LET} expressions, but reconstructing them with {\tt
directed-untranslate}.  These enhancements and others will be
incorporated into {\tt simplify-defun} when we add it, soon, to the
community books, located somewhere under {\tt books/kestrel}.

\section*{Acknowledgments}

This material is based upon work supported in part by DARPA under
Contract No. FA8750-15-C-0007.  We thank the referees, all of whom
gave very helpful feedback that we incorporated.

\bibliographystyle{eptcs}
\bibliography{paper}
\end{document}